\documentclass[epj]{svjour}
\usepackage{graphics}
\usepackage{amsmath}
\usepackage{amsfonts}
\usepackage{amssymb}

\usepackage{color}

\begin{document}
\title{Perturbations on the superconducting state of metallic nanoparticles: influence of geometry and impurities.}
\titlerunning{Perturbations of superconducting metallic nanoparticles}
\author{Mario Van Raemdonck\and Stijn De Baerdemacker  \and Dimitri Van Neck 
}                     
%
%
\institute{ Center for Molecular Modeling, Ghent University, Technologiepark 903, B-9052 Gent, Belgium\\
Department of Physics \& Astronomy, Ghent University, Proeftuinstraat 86, B-9000 Gent, Belgium}
\date{\today}
%
\abstract{
The pair condensation energy of a finite-size superconducting particle is studied as a function of two control parameters.  The first control parameter is the shape of the particle, and the second parameter is a position-dependent impurity introduced in the particle.  Whereas the former parameter is known to induce strong fluctuations in the condensation energy, the latter control parameter is found to be a more gentle probe of the pairing correlations.
} 
\maketitle
\section{Introduction}
In a seminal paper, P.\ W.\ Anderson \cite{anderson:1959} addressed the question of how small a metallic grain would have to be for that a superconducting state would cease to exist.  He argued that quantum confinement would force the single-particle (sp) spectrum to become discretely resolved.  The mean sp energy spacing will increase with decreasing size of the grain, until it becomes comparable to the superconducting gap in the bulk phase.  At that point, the bulk gap looses its significance as a clear gap between a single superconducting state and a continuum of excited states, and the superconducting phase would evaporate.   The single-electron transistor (SET) experiments of Ralph, Black and Tinkham \cite{ralph:1995} demonstrated that the energy spectrum of $nm$-scale Al particles is discretely resolved, and moreover, the spectrum was found to be dependent on number parity and externally applied magnetic fields  \cite{black:1996,ralph:1997}, establishing the persistence of pairing correlations at the nanoscale.   Bardeen, Cooper and Schrieffer (BCS) \cite{bardeen:1957} had identified an effective electron-electron pairing interaction as the driving force behind the superconducting state in bulk materials.  A key feature of BCS theory is that the superconducting ground state of a superconductor is modeled as a coherent condensation of Cooper pairs \cite{cooper:1956}.  While this approximation is essentially valid in the bulk limit, it is no longer sound for finite-size systems because inaccuracies induced by particle-number fluctuations become relatively large.  This opened a call for theoretical approaches in the canonical regime, such as Lanczos diagonalisation \cite{mastelloni:1998}, projected-BCS \cite{braun:1998}, Density Matrix Renormalization Group \cite{dukelsky:1999}, or the Richardson-Gaudin (RG) formalism \cite{sierra:2000}.  Richardson had shown, with\-in the context of pairing in nuclear structure physics \cite{dean:2003}, that the reduced, level-in\-de\-pendent, BCS Hamiltonian is exactly diagonalizable by means of a Bethe Ansatz product state, provided the RG variables, which occur as rapidities in the Ansatz, form a solution to the set of RG equations \cite{richardson:1963,richardson:1964a}.  Later, Gaudin decomposed the reduced BCS Hamiltonian into a complete set of commuting conserved charges, adding it to the class of  \emph{integrable} mo\-dels \cite{gaudin:1976}.  The main significance of these results is that, as long as a level-independent pair scattering term is considered, the BCS Hamiltonian can be solved exactly for a general sp spectrum, within polynomially scaling computing time.  Therefore, it is a practical tool for the investigation of pairing correlations in mesoscopic systems as a function of the sp spectrum.  First investigations were performed with a uniform sp energy spacing \cite{mastelloni:1998,braun:1998}, however, studies with randomly generated spectra showed an enhancement of pairing correlations by randomness \cite{smith:1996,sierra:2000}. This is related to the observation that pairing correlations are significantly stronger around the Fermi level, such that a random increase of the level density around the Fermi level will have a stronger impact on the mean pair correlations in a uniform sampling.  The result of this study triggers the question whether pairing correlations could be enhanced in a controlled fashion.  A sensible control parameter for the sp spectrum would be the shape and size of the nanoparticle. In a free-wave ''particle in a box'' picture, the geometric boundary conditions at the surface of the nanoparticle will fix the spectrum of the particles.   The variations in pairing correlations of a rectangularly shaped nanoparticle were investigated in this way as a function of the aspect ratio \cite{gladilin:2002}, and more recently, the shell structure in spherically shaped nanostructures has been assessed \cite{kresin:2006,garciagarcia:2011,croitoru:2011} in connection with the scanning tunneling experiments (STM) on deposited Pb \cite{brihuega:2011} and Sn \cite{bose:2010} nanoclusters.  The reduction from three to two dimensions, relevant for the description of pairing correlations in superconducting spherical coatings or multielectron bubbles in liquid helium, has also been investigated \cite{tempere:2005,gladilin:2006}.  The theoretical results in rectangular geometries showed a strongly volatile behavior of the pairing condensation as a function of the shape control parameter (see \emph{e.g.} Fig.\ 1 in \cite{gladilin:2002}), which is understood as a direct consequence of rapid fluctuations in the density of active sp levels around the Fermi level (also referred to as the Debeye window) \cite{bose:2010}.  

In the present contribution, we will focus deeper on how the pairing condensation energy varies as a function of an external control parameter.  The purpose of this work is to investigate whether there exists such a control parameter which is less prone to strong fluctuations and allows for a more controlled manipulation of the condensation energy.  Our calculations will be performed within the canonical RG formalism \cite{richardson:1963,richardson:1964a}, using a recently proposed pseudo-deformed quasispin algorithm \cite{debaerdemacker:2012b}.  In the next section, we recapitulate the necessary theoretical results of the RG formalism for metallic nanograins.  The following section is devoted to a scrutiny of the effect of the fluctuating level densities on the condensation energy for a small and easily fathomable system.  This section is divided into two parts.  In close parallel to the work of Gladilin \emph{et.\ al.\ }\cite{gladilin:2002}, we briefly discuss geometrical effects on the condensation energy in a first part.  In a second part, we introduce an impurity in otherwise clean nanograins.

For a good review on the developments in the field of superconducting metallic nanograins until 2001, we would like to refer the reader to the review paper of von~Delft and Ralph \cite{vondelft:2001}.
\section{Richardson-Gaudin}
The reduced BCS Hamiltonian is given by
\begin{equation}
\hat{H}=\sum_{i=1}^m \varepsilon_i \hat{n}_i + g\sum_{i,j=1}^m\hat{S}_i^\dag S_j,
\end{equation}
with the latin indices $\{i,j=1\dots m\}$ referring to a set of doubly-degenerate sp energies within the Debeye window around the Fermi level.  The number operator
\begin{equation}\label{rg:numberoperator}
\hat{n}_i = a^\dag_i a_i + a^\dag_{\bar{i}}a_{\bar{i}},
\end{equation}
counts the number of particles within a level $i$ and the pair scattering term is represented by the pair cre\-a\-tion/an\-ni\-hi\-la\-tion operators
\begin{equation}\label{rg:paircreationannihilation}
\hat{S}^\dag_{i}=a^\dag_ia^\dag_{\bar{i}},\quad \hat{S}_i=(\hat{S}_i^\dag)^\dag=a_{\bar{i}}a_i,
\end{equation}
with $a^\dag_i$ ($a_i$) the standard fermion creation (annihilation) operators.  The bar notation refers to the time-reversed partner of the corresponding operator.  The set of operators (\ref{rg:numberoperator}) and (\ref{rg:paircreationannihilation}) span an $su(2)$ quasispin algebra 
\begin{equation}
 [\hat{S}^0_i,\hat{S}^\dag_j]=\delta_{ij}S^\dag_i,\quad[\hat{S}^0_i,\hat{S}_j]=-\delta_{ij}S_i,\quad[\hat{S}^\dag_i,\hat{S}_j]=2\delta_{ij}S^0_i,
\end{equation}
with $\hat{S}^0_i=\frac{1}{2}(\hat{n}_i-1)$.  This algebra supports two different $su(2)$ representations, corresponding to unblocked (open) and blocked (pair-broken) levels.  In the present manuscript, only open levels will be considered.  Richardson's result \cite{richardson:1963,richardson:1964a} states that the reduced BCS Hamiltonian can be diagonalised exactly by means of a product state of generalised pairs, acting on the pair vacuum $|\theta\rangle$
\begin{equation}
 |\psi(\{x\})\rangle=\prod_{\alpha=1}^N\left(\sum_{i=1}^m\frac{\hat{S}^\dag_i}{2\varepsilon_i-x_\alpha}\right)|\theta\rangle,
\end{equation}
provided the set of RG variables $\{x\}=\{x_1,x_2,\dots x_N\}$, with $N$ the total number of pairs, form a solution of the set of non-linear RG equations
\begin{equation}
1+g\sum_{i=1}^m\frac{1}{2\varepsilon_i-x_\alpha}-2g\sum_{\beta\neq\alpha}\frac{1}{x_\beta-x_\alpha}=0,
\end{equation}
for $\alpha=1,2,\dots,N$.  Since we discard blocked levels, $N$ is taken as half the total number of particles.  Once the RG equations have been solved for $\{x\}$, the eigenstate energy of the Hamiltonian is given by
\begin{equation}
E=\sum_{\alpha=1}^N x_\alpha.
\end{equation}
It is convenient to introduce the concept of condensation energy, which is defined as the ground state energy $E$ of the system at a given pairing interaction $g$, corrected by the ground state energy of the system in the non-interacting limit $E_0$
\begin{equation}
 E_C=\langle\psi(g)|\hat{H}(g)|\psi(g)\rangle-\langle\psi(0)|\hat{H}(0)|\psi(0)\rangle
\end{equation}
Because the ground state of the non-interacting system corresponds to a simple filling of the sp levels until the Fermi energy, the condensation energy of the reduced BCS Hamiltonian reduces to
\begin{equation}\label{rg:condensationenergy}
 E_C=\sum_{\alpha=1}^Nx_\alpha-\sum_{i=1}^N2\varepsilon_i.
\end{equation}
The benefit of using the condensation energy over the ground state energy is that the former quantity corrects for global fluctuations in the sp energy, so it is a direct probe for pairing correlations.
\section{Perturbations}
We will employ a simplified ''particle in a box'' approach to study the effect of perturbations.  In this approach, it is assumed that the conductance electrons are completely delocalised from the atoms in the crystal, and move as free particles within a box, only confined by the boundaries.  Regarding the qualitative nature of our study, this approach satisfies our needs, however one should consider more sophisticated methods, such as Density Functional Theory \cite{hafner:2008}, if more realistic results are desired.
\subsection{Geometric perturbations}
\begin{figure}[!htb]
\includegraphics{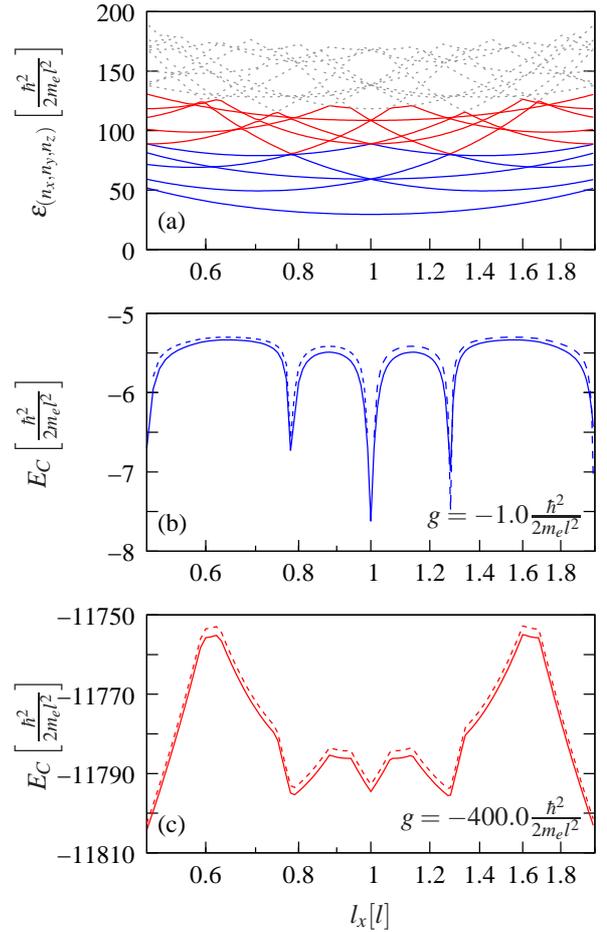}
\caption{The sp spectrum (a) of a rectangular box as a function of one of the dimensions $l_x$.  The lowest 10 levels within the active Debeye window are plotted in full lines, whereas the dotted lines depict the next 10 levels outside the Debeye window.  Figure (b) and (c) depict the condensation energy $E_C$ in full lines for a system of $N=5$ pairs in the $m=10$ active levels of Figure (a), for respectively the weak- and strong-coupling regime.  The dotted lines are approximations in respective regimes (eqs.\ (\ref{pert:geom:weakg}) and (\ref{pert:geom:strongg})).  Note that the $l_x$ axis is plotted in logarithmic scale to highlight the $l_x\leftrightarrow1/l_x$ symmetry.  Deviations of this symmetry are due to differences in resolution on the logarithmic scale.}\label{fig:balk}
\end{figure}

In a first part, we study and compare the condensation energy $E_C$ within a rectangular, cylindrical and spherical geometry.  The single-particle energies are taken as the solutions to the Schr\"odinger equation with the infinite-well sp potential $V(\vec{r})$ 
\begin{equation}
V(\vec{r})=\left\{\begin{array}{rl} 0,      & \forall \vec{r} \in\textrm{ the box}\\
                                    +\infty, & \forall \vec{r} \notin\textrm{ the box}\end{array}\right.\label{pert:geom:pot}
\end{equation}
The wavefunctions and corresponding eigenvalues for rectangular, cylindrical and spherical infinite wells can be found in introductory quantum mechanics textbooks \cite{bransden:2000}.  For rectangular geometries, the sp energies are given by
\begin{equation}\label{pert:geom:boxsp}
 \varepsilon_{(n_x,n_y,n_z)}=\frac{\hbar^2\pi^2}{2m_e}\left(\frac{n_x^2}{l_x^2}+\frac{n_y^2}{l_y^2}+\frac{n_z^2}{l_z^2}\right),
\end{equation}
with $(n_x,n_y,n_z)\in\mathbb{N}_0^3$ the set of quantum numbers, $m_e$ the effective mass of the electron, and $(l_x,l_y,l_z)$ the dimensions of the rectangular box.  We will define a length scale $l$, such that all lengths will be given relative to $l$, and energies relative to $\frac{\hbar^2}{2m_el^2}$.  The sp spectrum (\ref{pert:geom:boxsp}) is given in Figure \ref{fig:balk}(a) as a function of $l_x$, with $l_z=l$ kept as a constant, and $l_y$ defined as such that the volume of the rectangular box also remains a constant $l_xl_yl_z=l^3$. 
In addition, Figures \ref{fig:balk}(b) and \ref{fig:balk}(c) show the condensation energy for a system of $N=5$ pairs living in the $m=10$ lowest sp levels of the rectangular box with respectively a weak- ($g=-1.0[\hbar^2/2m_el^2]$) and strong ($g=-400.0[\hbar^2/2m_el^2]$) pairing interaction.  Figure \ref{fig:balk}(a) not only depicts the 10 active sp levels, but also the next 10 levels outside of the Debeye window (in dotted lines) to illustrate how the sp levels enter and leave the Debeye window as a function of $l_x$.  It can be seen that the steep exits and enterings of the sp levels into the Debeye window give rise to strong fluctuations in the sp densities, especially higher up in the spectrum.  This has an effect on the condensation energies, as can be inferred from Figures \ref{fig:balk}(b) and \ref{fig:balk}(c).  

Before discussing the numerical results, it is worthwhile to distill the general features in both regimes of the pairing strength using perturbative techniques.  For the weak-coupling regime, standard 2nd order perturbation theory \cite{bransden:2000} is used to calculate the condensation energy
\begin{equation}\label{pert:geom:weakg}
 \lim_{g\rightarrow0}E_C=Ng+g^2\sum_{a=1}^{k_F}\sum_{b=k_F+1}^m\frac{1}{2\varepsilon_a-2\varepsilon_b}+\mathcal{O}(g^3),
\end{equation}
with $k_F$ the Fermi level index, and the dummy indices $a$ and $b$ running over respectively occupied and unoccupied sp levels in the non-interacting limit.   For the strong-coupling regime, an approximate expression can be derived for the condensation energy using the RG equations \cite{debaerdemacker:2012b,yuzbashyan:2003}
\begin{equation}\label{pert:geom:strongg}
\lim_{g\rightarrow\infty}E_C=gN(m-N+1)+N(\langle2\varepsilon\rangle-\langle2\varepsilon\rangle_F)+\mathcal{O}(\tfrac{1}{g}),
\end{equation}
with $\langle2\varepsilon\rangle=\frac{1}{m}\sum_{i=1}^m2\varepsilon_i$ the mean pair sp energy and $\langle2\varepsilon\rangle_F=\frac{1}{N}\sum_{i=1}^{k_F}2\varepsilon_i$ the mean pair sp energy up to the Fermi level.  Close inspection of the functional behavior of expressions (\ref{pert:geom:weakg}) and (\ref{pert:geom:strongg}) with respect to the sp spectrum gives away the gross features in the corresponding regime.  In the strong-coupling regime, the condensation energy is dependent on $\langle2\varepsilon\rangle$ and $\langle2\varepsilon\rangle_F$.  Therefore, the contributions of the sp levels on the condensation energy only depend on the position of the level with respect to the Fermi level, \emph{i.e.} levels beneath the Fermi level contribute negatively with a weight factor $(\frac{N}{m}-1)$, and levels above the Fermi level contribute positively with weight factor $(\frac{N}{m})$.  There is no direct dependency of local sp-level densities on the condensation energy, rather an indirect dependency entering via the mean sp energy above and below the Fermi level.  For instance, it can be seen from Figure \ref{fig:balk}(c) that the local increase in sp-level density around the Fermi level in the vicinity of $l_x\sim0.8$, $1.0$ and $1.25$ induces an increase of $\langle2\varepsilon\rangle_F$, and therefore contributes to the pairing condensation.  Similarly, the local increase in sp-level density at the top of the Debeye window around $l_x\sim0.6$ and $1.6$ contributes positively to $\langle2\varepsilon\rangle$ and therefore decreases the pairing correlations.  The situation is different for the weak-coupling regime, where local density fluctuations around the Fermi level contribute more strongly than those away from the Fermi level.  This can be verified in Figure \ref{fig:balk}(b), where, in contrast to the strong-coupling regime, the local density increase at $l_x\sim0.6$ and $1.6$ does not influence the condensation energy, whereas the the local density increase at $l_x\sim0.8$, $1.0$ and $1.25$ considerably enhances the pairing correlations. 

The previous discussion highlights the importance of local sp-level density fluctuations entering directly, or indirectly into the condensation energy.  Therefore, it would be desirable to have a more manageable control parameter at hand for the sp-energy levels.  As is clear from Ref.\ \cite{gladilin:2002} and Figure \ref{fig:balk}(a), rectangular geometries are very prone to local density fluctuations and will consequently remain hard to control.  For this reason, we performed a similar study of the condensation energy within a cylindrical and spherical configuration.  Without going into much detail, the conclusions of these studies agreed well with the results from the previous discussion.  The sp spectrum of free particle waves, bounded within a cylinder with radius $\rho_0$ and height $l_z$ is given by
\begin{equation}\label{pert:geom:cylindersp}
 \varepsilon_{(n_\rho,n_\phi,n_z)}=\frac{\hbar^2}{2m_e}\left(\frac{\alpha^2_{|n_\phi|n_\rho}}{\rho_0^2}+\frac{n_z^2\pi^2}{l_z^2}\right),
\end{equation}
with $(n_\rho,n_\phi,n_z)\in\mathbb{N}\times\mathbb{Z}\times\mathbb{N}$, and $\alpha_{|n_\phi|n_\rho}$ the $n_\rho$th root of the Bessel function $J_{|n_\phi|}(\alpha)$.  It is clear that the spectrum (\ref{pert:geom:cylindersp}) and its density fluctuations has a qualitatively similar dependency on the control parameter $l_z$ as eq.\ (\ref{pert:geom:boxsp}), when volume conservation is imposed $\pi\rho_0^2l_z=l^3$.  Therefore similar conclusions could be drawn for cylindrical as for rectangular geometries.  

Unfortunately, the sphere has no shape control parameter if volume conservation is applied $4\pi\rho_0^3=l^3$.  However, one can notice a gain in condensation energy in comparison with a cube and cylinder with the same dimensions (see Table \ref{table:condensationenergy}).  
\begin{table}[!htb]
\begin{center}
\caption{The condensation energy $E_C$ (\ref{rg:condensationenergy}) of a cube, cylinder and sphere with volume $l^3$ for 3 different values of the pairing interaction strength, corresponding to a weak-, intermediate- and strong-coupling regime.  The radius $\rho_0$ of the cylinder is fixed as such that the height $l_z=l$.  All calculations have been performed with $N=6$ pairs in the $m=12$ first sp levels.  Energies are given in units $[\hbar^2/2m_el^2]$}
\label{table:condensationenergy}       
\begin{tabular}{r|rrr}
\hline\noalign{\smallskip}
   $g$  &     cube & cylinder &   sphere \\
\noalign{\smallskip}\hline\noalign{\smallskip}
 -1.000 &   -8.851 &   -6.850 &  -13.067 \\
-10.000 & -222.254 & -226.183 & -246.037 \\
-20.000 & -600.819 & -608.382 & -626.128 \\
\noalign{\smallskip}\hline
\end{tabular}
\end{center}
\end{table}
This is nicely understood via symmetry considerations; the sphere is more symmetric than the cube, and will therefore exhibit more degeneracies in the sp spectrum, leading to an enhancement of the pair correlations.  Analogously, a cylinder is more symmetric than a cube (2D rotational vs dihedral symmetry), which generally translates into an enhancement for the pairing correlations, as illustrated in Table \ref{table:condensationenergy}.  The symmetry argument  can be tested by breaking the symmetry of the sphere or cylinder to spheroidal shapes.  Although such a study is of relevance for the experiments on spherical nanodroplets \cite{brihuega:2011,bose:2010}, we do not expect significant qualitative differences from our study with rectangular grains and leave this subject for future investigations.   

The conclusion of the present subsection is that the condensation energy of a rectangular (and cylindrical) na\-no\-grain is highly sensitive to the fluctuations in the sp-level densities, and that these fluctuations are rather strong as a function of the shape control parameter.  With this respect, it would be interesting to find a more gentle control parameter such that the pairing correlations can be probed in a more controlable fashion.  In the next subsection, we introduce impurities for this particular purpose. 

\subsection{Impurities}
\begin{figure}[!htb]
\includegraphics{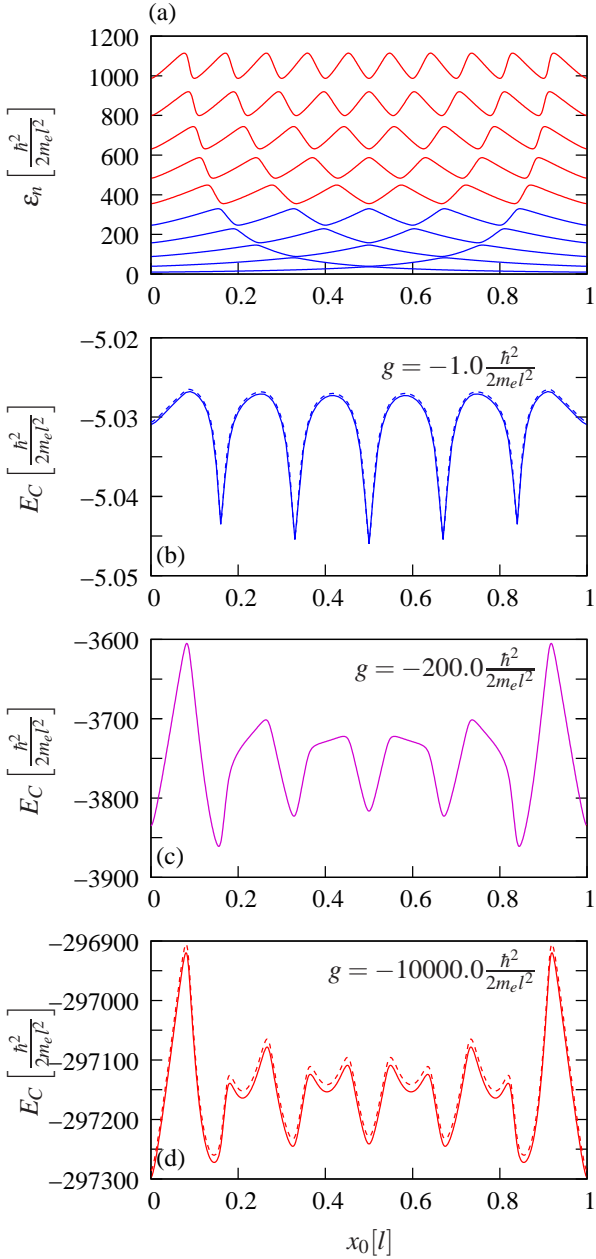}
\caption{In panel (a), the sp spectrum (\ref{pert:imp:sp}) of a 1D particle-in-a-box with a $\delta(x-x_0)$ impurity at $x_0$ is depicted. Panels (b)-(d) show the condensation energy as a function of $x_0$, for the weak- (b), intermediate- (c), and strong-coupling (d) regime.  For the weak-, and strong-coupling regime, the approximative predictions, given in respectively Eqs.\ (\ref{pert:geom:weakg}) and (\ref{pert:geom:strongg}), are plotted in dashed lines.}\label{fig:delta}
\end{figure}

From a ''particle in a box'' perspective, an impurity can be modeled by means of an ''obstacle'' in the otherwise constant potential of the box.  Let this obstacle be a Dirac $\delta(\vec{r})$ potential.  For a 1D system, the potential in the Schr\"odinger equation becomes
\begin{equation}
V(x)=\left\{\begin{array}{ll} v_0l\delta(x-x_0)  & 0<x<l \\
                               \infty           & x\leq0\textrm{ and }x\geq l\end{array}\right.\label{pert:imp:delta}
\end{equation}
with $0<x_0<l$ and $v_0$ a weighted strength parameter of the impurity which can be either positive or negative, depending on whether the impurity is considered repulsive or attractive.  The solution to the Schr\"odinger equation can be found by solving the following transcendental equation
\begin{equation}\label{pert:imp:transeq}
\tfrac{2m_el^2}{\hbar^2}v_0\sin(k[l-x_0])\sin(kx_0)+kl\sin(kl)=0,
\end{equation}
for $k\in\mathbb{R}$, leading to the sp spectrum
\begin{equation}\label{pert:imp:sp}
\varepsilon_n=\frac{\hbar^2k_n^2}{2m_e}.
\end{equation}
If $v_0<0$, there may also exist a negative energy state, which is the solution of Eq.\ (\ref{pert:imp:transeq}), with the substitution $ik\rightarrow\kappa$
\begin{equation}
\tfrac{2m_el^2}{\hbar^2}v_0\sinh(\kappa[l-x_0])\sinh(\kappa x_0)+\kappa l\sinh(\kappa l)=0.
\end{equation}
In the remainder of the paper, we will only deal with repulsive impurities ($v_0>0$).  The transcendental equation (\ref{pert:imp:transeq}) has a few remarkable symmetries.  For instance, it can be verified that for $x_0=\frac{l}{p}$ with $p\in\mathbb{N}_0$, $k=\frac{q\pi}{l}$ is always a solution of (\ref{pert:imp:transeq}) independent from $v_0$, as long as $q$ is a multiple of $p$.  This feature explains the quasi-periodic structure as a function of $x_0$ in the sp spectrum, which is plotted in Figure~\ref{fig:delta}(a) for $v_0=100.0[\hbar^2/2m_el^2]$.  
It can be seen that each level has its own frequency modulation: the first level undergoes one full quasi-period oscillation, the second level a double quasi-period oscillation, and so on.  Therefore, by probing the oscillations in the condensation energy, it can be inferred which sp levels contribute strongly to the final structure.  We can recall from the previous discussion that the condensation energy is more sensitive to local sp-level density fluctuations around the Fermi level in the weak-coupling regime, whereas global fluctuations contribute more in the strong-coupling regime.  This can be observed in Figures \ref{fig:delta}(b)-(d), where the condensation energy is plotted for a system of $N=5$ pairs, living in the first $m=10$ sp levels of Figure \ref{fig:delta}(a), with $g=-1.0$, $-200.0$, and $-100 000.0$ in units $[\hbar^2/2m_el^2]$.  Taking into account that the mean sp-energy spacing is approximately $100.0(\hbar^2/2m_el^2)$, these interaction strengths correspond respectively to the weak, intermediate, and strong-coupling regime.  In the weak-coupling limit (Figure \ref{fig:delta}(b)), the condensation energy displays 5 peaks of enhanced pairing correlations, corresponding to the 5 quasi periods of the Fermi level.  On the other side, the 10 quasi periods at the top of the Debeye window are visible in the condensation energy of the strong-coupling limit (Figure \ref{fig:delta}(d)).  The intermediate regime (Figure \ref{fig:delta}(c)) displays only 5 quasi periods, but it can be inferred from the shape of the modulations, that the signature of the top level is already present.

The reason why the impurity is a much more gentle control parameter than the shape of a nanograin can be related to the relative impact of the control parameter on the sp spectrum.  Whereas  altering the size of the nanograin has a large relative effect on the available space of the particles-in-a-box, adding a $\delta(x)$ only perturbs the particles marginally.  The question is now whether $\delta(\vec{r})$ perturbations are not becoming too weak when going to higher dimensions.  In order to study this, we performed some exploratory calculations of the condensation energy with one $\delta(\vec{r}-\vec{r}_0)$ on a line (1D), in a square (2D), and in a cube (3D).  The number of levels $m=10$ and pairs $N=5$ were chosen equal for each dimension, as well as the strength of the impurity  $v_0=100.0[\hbar^2/2m_el^2]$, and the pairing strength $g$.  These preliminary calculations pointed out that the condensation energy was enhanced with approximately 20\% and 25\% for the 2D and 3D systems respectively compared to the 1D case, whereas the relative fluctuations in the condensation energy decreased with 75\% and 85\%.  These numbers hint at a possible survival of impurity induced fluctuations in the condensation energy in higher dimensions, but further investigations are required.  More in particular, given the strong influence of the geometry on the condensation energy of the particle, it is unclear whether the gentle impurity-induced perturbations will be observable against the large geometric fluctuations one could encounter by picking different samples in a realistic experimental setting.  

\begin{figure}[!htb]
\includegraphics{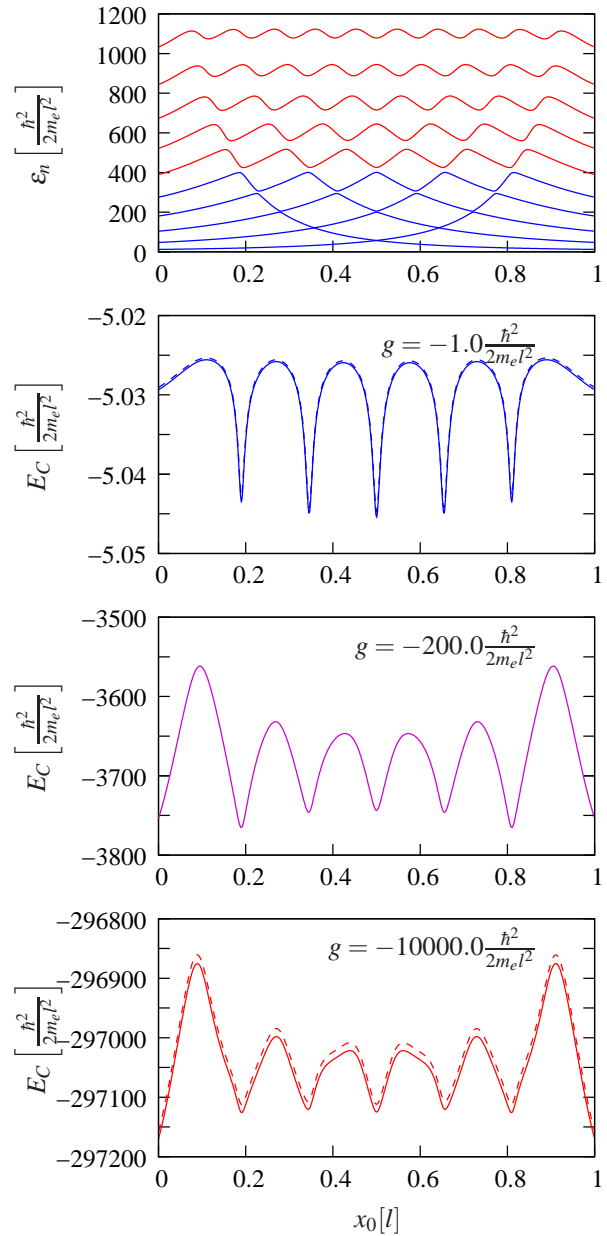}
\caption{In panel (a), the sp spectrum (\ref{pert:imp:sp}) of a 1D particle-in-a-box with a $\exp{(-k|x-x_0|)}$ impurity at $x_0$ and $k=20/l$ is depicted. Panels (b)-(d) show the condensation energy as a function of $x_0$, for the weak- (b), intermediate- (c), and strong-coupling (d) regime.  For the weak-, and strong-coupling regime, the approximative predictions, given in respectively Eqs.\ (\ref{pert:geom:weakg}) and (\ref{pert:geom:strongg}), are plotted in dashed lines.}\label{fig:deltascreened}
\end{figure}

The $\delta(x)$ potential has zero-range character, in contrast to the spatial finite-range nature of realistic impurities in 1D systems, which may become relatively large for nano-sized systems \cite{reyes:1998}.  Therefore, we have selected the following potential
\begin{equation}
V(x)=\left\{\begin{array}{ll} v_0\frac{kl}{2}\exp(-k|x-x_0|)  & 0<x<l \\
                               \infty           & x\leq0\textrm{ and }x\geq l\end{array}\right.,\label{pert:imp:exp}
\end{equation}
to investigate the effect of the spatial extent of the impurity on the condensation energy of the 1D system.  Besides being a solution of the Helmholtz equation for screened Coulomb potentials of a point-like charged particle in 1D, the potential (\ref{pert:imp:exp}) acts as a distribution, including the $\delta(x-x_0)$ potential (\ref{pert:imp:delta}) and unperturbed system in the $k\rightarrow\infty$ and $k\rightarrow0$ limit respectively.  Therefore, the parameter $1/k$ is a control parameter for the spatial extent of the impurity.  We have carried out the same calculation as in Figure \ref{fig:delta} with the same values for $v_0$ and $g$, but for different values of $k$ ranging from very large ($\delta(x)$-like) to very small (unperturbed-like) values.  The results for $k=20/l$ are depicted in Figure \ref{fig:deltascreened}.
The potential (\ref{pert:imp:exp}) will deviate from the $\delta(x-x_0)$ potential as $k$ decreases, so the typical modulations in the sp spectrum and condensation energy of the $\delta(x-x_0)$ case are expected to gradually evaporate as the potential (\ref{pert:imp:exp}) broadens.  For the sp spectrum, it was observed that the modulations were more suppressed for the high-lying states, compared to the low-lying states.  Because the normalization of the potentials (\ref{pert:imp:exp}) and (\ref{pert:imp:delta}) has been chosen equal, the potential (\ref{pert:imp:exp}) has a finite height $V_{\max}=v_0lk/2$, in contrast with the infinite height of the $\delta(x)$ potential.  As a result, the higher-lying excitation sp levels will be less affected by the impurity than the lower-lying sp levels (see Figure \ref{fig:deltascreened}(a) with $V_{\max}=1000[\hbar^2/2m_el^2]$).  This has an effect on the modulations of the condensation energy in the strong-interaction limit.  Because the modulations of the condensation energy in the strong-interaction limit depends approximately on  the relative weighting of the sp levels above and beneath the Fermi level $\varepsilon_F$, the fingerprints of the higher sp levels will gradually disappear as $k$ decreases.  The value $k=20/l$ has been chosen for Figure \ref{fig:deltascreened} because this is the point where the higher sp level modulations start to (visually) disappear from the condensation energy in the strong-interaction limit (see Figure \ref{fig:deltascreened}(d)).  The condensation energy in the weak-interacting limit is only dependent on the modulations around the Fermi level (see eq.\ (\ref{pert:geom:weakg})).  So, we will observe modulations in the condensation energy as long as the levels around the Fermi level are affected by the impurity.  Again, this is strongly dependent on the relative position of the Fermi level with respect to the height and strength of the impurity potential.  In the limit of $k\rightarrow0$, all impurity induced structure will be lost.

\begin{figure}[!htb]
\includegraphics{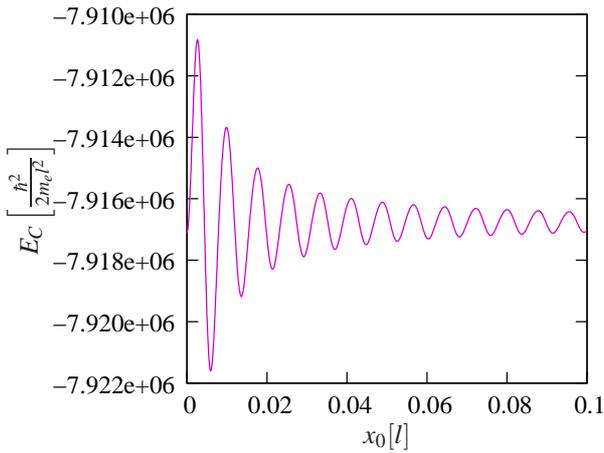}
\caption{The condensation energy of a system consisting of $N=128$ pair in $m=256$ levels, with a $\delta(x-x_0)$ impurity, as a function of the position $x_0$.  The strength of the impurity has been chosen as $v_0=100.0[\hbar^2/2m_el^2]$ and the pairing strength is $g=-2000.0[\hbar^2/2m_el^2]$.  For graphical reasons, the condensation energy is only given in the interval $x_0\in[0,0.1]$.}\label{fig:huge}
\end{figure}

Finally, we also performed a calculation of the condensation energy for $N=128$ pairs living in $m=256$ levels of the 1D system with a $\delta(x-x_0)$ impurity.  From a physics point of view, this particular size of system corresponds to a realistic number of active electron pairs within the Debeye window of a nanograin.  The Richardson-Gaudin proves particularly useful in this particular situation because the size of the Hilbert space ($\dim\sim5.7\times10^{75}$) is far beyond the capabilities of standard diagonalisation approaches.  The result of the calculation is presented in Figure \ref{fig:huge}.
The effect of the larger number of particles and levels on the condensation energy is immediately visible in the modulation, which has increased to 128 quasi periods, corresponding to the number of quasi periods of the Fermi level.  Therefore, our analysis for the smaller system appears to be valid in larger systems as well.  

\section{Conclusions}
In conclusion, we have studied the effect of two qualitatively different control parameters on the pair condensation energy of a finite-size superconducting particle.  The control parameters enter into the system via the single-particle spectrum, which is based on a straightforward particle-in-a-box principle.  The first control parameter is the shape of the particle, which induces strong fluctuations in the single-particle level densities, precipitating into the condensation energy.  By means of perturbation theory, it was found that the condensation energy in the weak-coupling regime is mainly dependent on local single-particle level density fluctuations, whereas the strong-coup\-ling regime is also affected by global level density fluctuations.  Introducing impurities, as a second control parameter, proved to be a more gradual probe for pairing correlations.  An impurity gives a unique quasi-periodic structure to each single particle level as a function of the position of the impurity, such that it becomes possible to weigh the contributions of the single-particle levels to the condensation energy by investigating the frequency of oscillations.  Regarding the schematic nature of this study, possibilities to bring these results towards more realistic 2D and 3D system is briefly touched upon.

We acknowledge interesting conversations with Kris Van Houcke, Kris Heyde (Ghent University) and stimulating discussions with Margriet Van Bael (KULeuven) during the International Symposium on Small Particles and Inorganic Clusters XVI.  SDB is a ''FWO-Vlaanderen'' postdoctoral fellow.

%
%
%
%
\bibliographystyle{spphys}       
\bibliography{vanraemdonck_pertscgs} 
%

\end{document}